\documentclass[preprint,secnumarabic,amssymb, nobibnotes, aps, prd]{revtex4-1}

\usepackage{graphicx}

\begin{document}


\title{Fixed-scale approach to finite-temperature lattice QCD with shifted boundaries} 


\author{Takashi Umeda}
\email{tumeda@hiroshima-u.ac.jp}
\affiliation{Graduate School of Education, Hiroshima University, Hiroshima 739-8524, Japan}
\date{\today}

\begin{abstract}
We study the thermodynamics of the SU(3) gauge theory using the fixed-scale approach with shifted boundary conditions.
The fixed-scale approach can reduce the numerical cost of the zero-temperature part in the equation of state calculations, while the number of possible temperatures is limited by the integer $N_t$, which represents the temporal lattice extent. The shifted boundary conditions can overcome such a limitation while retaining the advantages of the fixed-scale approach. Therefore, our approach enables the investigation of not only the equation of state in detail, but also the calculation of the critical temperature with increased precision even with the fixed-scale approach. We also confirm numerically that the boundary conditions suppress the lattice artifact of the equation of state, which has been confirmed in the non-interacting limit.
\end{abstract}

\maketitle

\section{Introduction}
Lattice QCD is the sole method developed thus far to calculate QCD thermodynamics nonperturbatively at intermediate temperatures. The calculation of bulk thermodynamic observables has been established on the lattice, and the approach has uncovered various properties of the quark-gluon-plasma (QGP), such as its transition temperature, order of the transition, and the equation of state (EOS). The lattice results are indispensable to understand the QGP created in heavy-ion collision experiments as inputs of the hydrodynamical description of QGP space-time evolution \cite{Hirano:2008hy}. In the last decade, the dynamical quark effects in QCD thermodynamics on lattices have been studied in detail. 
In particular, $2+1$ flavor QCD with the physical quark masses for degenerate up and down plus strange quarks 
have been realized on the lattice by using staggered-type quarks \cite{Borsanyi:2013bia, Bazavov:2014noa}.

However, in principle, the use of staggered-type quarks is not applicable
to the case that the number of flavors is not a multiple of four.
In such case, calculations with theoretical sound quarks, e.g., Wilson quarks, are utilized to check the validity of results by using the staggered-type quarks.
However, with such sound quarks, the calculation of the physical light quark masses requires huge computational resources. Therefore, we require more efficient approaches to study QCD thermodynamics on the lattice.

To this end, we have previously proposed the fixed-scale approach \cite{Umeda:2008bd}, in which temperature is varied by varying the temporal lattice size $N_t$ at a fixed lattice spacing $a$, while in the conventional approach, $a$ is varied for a fixed $N_t$. 
The fixed-scale approach provides various advantages to calculate the EOS on lattices. 
One of the advantages involves the numerical cost for zero-temperature simulations, which accounts for a large portion of the total amount of EOS calculation. We first remark that thermodynamic quantities sometimes require subtraction of the zero-temperature contribution, which has to be performed for each coupling parameter.
Second, in order to keep physical conditions except for temperature, we have to identify a line of constant physics in the coupling parameter space. This limitation forms a heavy computational burden for full QCD simulations, in which the quark mass parameters should be determined such that, for example, the ratios of hadron masses remain constant.
Third, the beta functions to obtain the EOS are necessary for each set of coupling parameters. These are also determined by zero-temperature simulations. 

In the fixed-scale approach, a series of temperatures is given with a single lattice scale, i.e., a set of coupling parameters. Therefore, the zero-temperature subtraction and the beta-functions are common for the temperatures, and the condition for the line of constant physics is automatically satisfied. 
These advantages offered by the fixed-scale method yielded the first result regarding the EOS for $2+1$ flavor QCD with nonperturbatively improved Wilson quarks \cite{Umeda:2012er}. Furthermore, some groups have adopted this approach to study the EOS using the smeared Wilson \cite{Borsanyi:2012uq} and the twisted mass quarks \cite{Burger:2013hia}.

However, a disadvantage of the fixed-scale approach is the limited number of temperatures, which is restricted by the $N_t$ of integers.
Furthermore, the approach becomes complicated with the use of an even--odd algorithm that is often adopted to generate QCD gauge configurations.
Although we can assume multiple sets of coupling parameters, this rather negates the advantage of the fixed-scale approach.

In this backdrop, Giusti and Meyer proposed the shifted boundary condition as a method to calculate the thermodynamic quantities, e.g., entropy density \cite{Giusti:2010bb}. 
In the method, the shifted boundary conditions are introduced to evaluate the momentum distribution function that is naturally expressed by the path-integral with the shifted boundary conditions. The cumulants of the momentum distribution functions enable us to calculate the thermodynamic quantities by means of the Word identities \cite{Giusti:2011kt}. With the new method, for example, the entropy density is obtained numerically in the SU(3) gauge theory on the lattice \cite{Giusti:2010bb, Giusti:2014ila}.

As a by-product of the shifted boundary conditions, a fine temperature scan is possible as described in the following.
Here, we consider relativistic thermal field theories with the shifted boundaries, which are defined as the following conditions for the field $\phi(\vec{x},t)$,
\begin{eqnarray}
\phi(\vec{x}, L_0)=\pm \phi(\vec{x}+\vec{s},0).
\end{eqnarray}
Here $\vec{s}$ denotes a spatial shift vector, the $+(-)$ sign indicates bosonic (fermionic) fields, and $L_0$ denotes the temporal extent (inverse temperature). 
Due to the underlying Lorentz symmetry of the theories,
in the thermodynamic limit, the invariance of the dynamics under the SO(4) group implies that the free energy density $f(L_0; \vec{s})$ satisfies the relation \cite{Giusti:2012yj},
\begin{eqnarray}
f(L_0, \vec{s}) = f(\sqrt{L_0^2+\vec{s}^2},\vec{0}).
\end{eqnarray}

Therefore, the thermodynamic observables of the theories depend only on the combination $\sqrt{L_0^2+\vec{s}^2}$, which is equivalent to the inverse temperature of the system.
The expectation value of the observables at the temperature of $(L_0^2 + \vec{s}^2)^{-1/2}$ can be also obtained with the simulation at the temporal extent $L_0$ with the shifted boundary of the vector $\vec{s}$. 
Therefore, the number of possible temperature values is largely increased by varying the shift vectors in addition to the lattice temporal extent $N_t$ even in the fixed-scale approach.
Since the bare lattice parameters are common, the lattice scale is also fixed; in other words, the advantages of the fixed-scale approach hold provided the shifted boundary conditions are introduced.

In this study we test this idea in the SU(3) gauge theory on calculations of the trace anomaly and the critical temperature.
Our numerical setup is described in Sect. \ref{sec:setup}.
The trace anomaly and the critical temperature are discussed in Sect. \ref{sec:result}.
Our remarks on the beta function are presented in Sect. \ref{sec:beta}, and we summarize and conclude this study in Sect. \ref{sec:conclusion}.

\section{Lattice setup}
\label{sec:setup}
The calculations are performed in the SU(3) gauge theory. 
The standard plaquette gauge action is defined as

\begin{eqnarray}
S_g &=& 
6N_s^3N_t \beta P \\
P&\equiv &\frac{1}{6N_s^3N_t}\sum_{\vec{x},t}\sum_{\mu\neq\nu=1}^4
\left[ 1-\frac{1}{3}\mbox{Re}\mbox{Tr} U_{\mu\nu}(\vec{x},t) \right]
\end{eqnarray}

where $U_{\mu\nu}(\vec{x},t)$ denotes the product of the link variables $U_\mu(\vec{x},t)$ along a plaquette in the $\mu-\nu$ plane. We adopt the periodic boundary condition along the spatial directions and the shifted boundary condition along the temporal direction with the shift vector $\vec{s}$,

\begin{eqnarray}
U_4(\vec{x},N_t) = U_4(\vec{x}+\vec{s},0).
\end{eqnarray}

In this study, we adopt a bare lattice gauge coupling given by $\beta=6/g^2=6.0$, which is often adopted in such studies, and it is thus well-studied at finite temperatures also. The corresponding lattice spacing is about $0.1$ fm, which value is determined from the Sommer scale with $r_0=0.5$ fm. The primary calculations are performed with $N_s^3=32^3$ volume lattices, whose spatial size is about (3fm)$^3$ at each temperature with $N_t=$ 3--9. 
In the SU(3) gauge theory for the above parameter values, the critical temperature is located around the temperature at $N_t=7$ (with $\vec{s}=\vec{0}$) \cite{Umeda:2008bd}.
The Zero-temperature observables are measured on the symmetric lattice size of $32^4$.
The zero- and finite-temperature gauge configurations are generated by 
the pseudo-heat-bath update algorithm with over-relaxation. 
The shifted boundary conditions are incorporated into the pseudo-heat-bath algorithm by modifying only the staple constructions at the temporal boundary. 

The shift vectors adopted in our calculations are listed in Tab. \ref{tab:shift}.
Since our pseudo-heat-bath update code utilizes even--odd labeling in the spatial link variables to be optimized in vector machine computations, the shift vectors at the boundary are restricted to even shift vectors, which ensures even--odd labeling at the boundary. Our update code is able to generate configurations with the temporal extent of odd numbers in addition to ordinary even numbers. 
The system temperature is defined by $T=1/\sqrt{a^2N_t^2 + \vec{s}^2}$, whose inverses in lattice units are listed in Tab. \ref{tab:shift}.

\begin{table}[tb]
\begin{tabular}{|ccc|ccccccc|}
\hline
 & $\vec{s}$ & & &&& $N_t$ &&& \\
$s_1$ & $s_2$ & $s_3$ & 9 & 8 & 7 & 6 & 5 & 4 & 3 \\
\hline
0 & 0 & 0 & 9.00 & 8.00 & 7.00 & 6.00 & 5.00 & 4.00 & 3.00 \\
1 & 1 & 0 &  -   & 8.12 & 7.14 & 6.16 & 5.20 & 4.24 & 3.32 \\
2 & 0 & 0 &  -   & 8.25 & 7.28 & 6.32 & 5.39 & 4.47 & 3.61 \\
2 & 1 & 1 &  -   & 8.37 & 7.42 & 6.48 & 5.57 & 4.69 & 3.87 \\
2 & 2 & 2 &  -   & 8.49 & 7.55 & 6.63 & 5.74 & 4.90 & 4.12 \\
3 & 1 & 0 &  -   & 8.60 & 7.68 & 6.78 & 5.92 & 5.10 & 4.36 \\
2 & 2 & 2 &  -   & 8.72 & 7.81 & 6.93 & 6.08 & 5.29 & 4.58 \\
3 & 2 & 1 &  -   & 8.83 & 7.94 & 7.07 & 6.24 & 5.48 & 4.80 \\
4 & 0 & 0 &  -   & 8.94 & 8.06 & 7.21 & 6.40 & 5.66 & 5.00 \\
3 & 3 & 0 &  -   & 9.06 & 8.19 & 7.35 & 6.56 & 5.83 &  -   \\
4 & 1 & 1 &  -   &  -   &  -   & 7.35 & 6.56 & 5.83 &  -   \\
4 & 2 & 0 &  -   &  -   &  -   & 7.48 & 6.71 & 6.00 &  -   \\
3 & 3 & 2 &  -   &  -   &  -   & 7.62 & 6.86 & 6.16 &  -   \\
4 & 2 & 2 &  -   &  -   &  -   & 7.75 & 7.00 & 6.32 &  -   \\
4 & 3 & 1 &  -   &  -   &  -   & 7.87 & 7.14 &  -   &  -   \\
5 & 1 & 0 &  -   &  -   &  -   & 7.87 & 7.14 &  -   &  -   \\
5 & 2 & 1 &  -   &  -   &  -   & 8.12 & 7.42 &  -   &  -   \\
4 & 4 & 0 &  -   &  -   &  -   & 8.25 & 7.55 &  -   &  -   \\
4 & 3 & 3 &  -   &  -   &  -   & 8.37 &  -   &  -   &  -   \\
\hline
\end{tabular}
\caption{Inverse temperatures at each $N_t$ with boundary shifts $\vec{s}$. The $s_1$, $s_2$, and $s_3$ values correspond to the components of the shift vector $\vec{s}$. The numbers below $N_t$ at the top of the table indicate the lattice temporal extent $N_t$. The numbers written to the accuracy of two decimal places represent the inverse temperatures at each $N_t$. All numbers are in lattice units.}
\label{tab:shift}
\end{table}

We generated finite temperature configurations up to approximately 200,000 pseudo-heat-bath sweeps after 10,000 thermalization sweeps at each value of $N_t$ and $\vec{s}$. For zero temperature, the configurations were generated up to approximately 30,000 sweeps after 10,000 sweeps of thermalization.
Error analyses were performed using by the jackknife method unless otherwise stated.
The number of sweeps in a jackknife-bin for finite temperature was 2,000 sweeps, that is sufficient even near the critical temperature. That for zero temperature was 500 sweeps.

\section{Lattice results}
\label{sec:result}
\begin{figure}[tb]
\resizebox{75mm}{58mm}{
\includegraphics{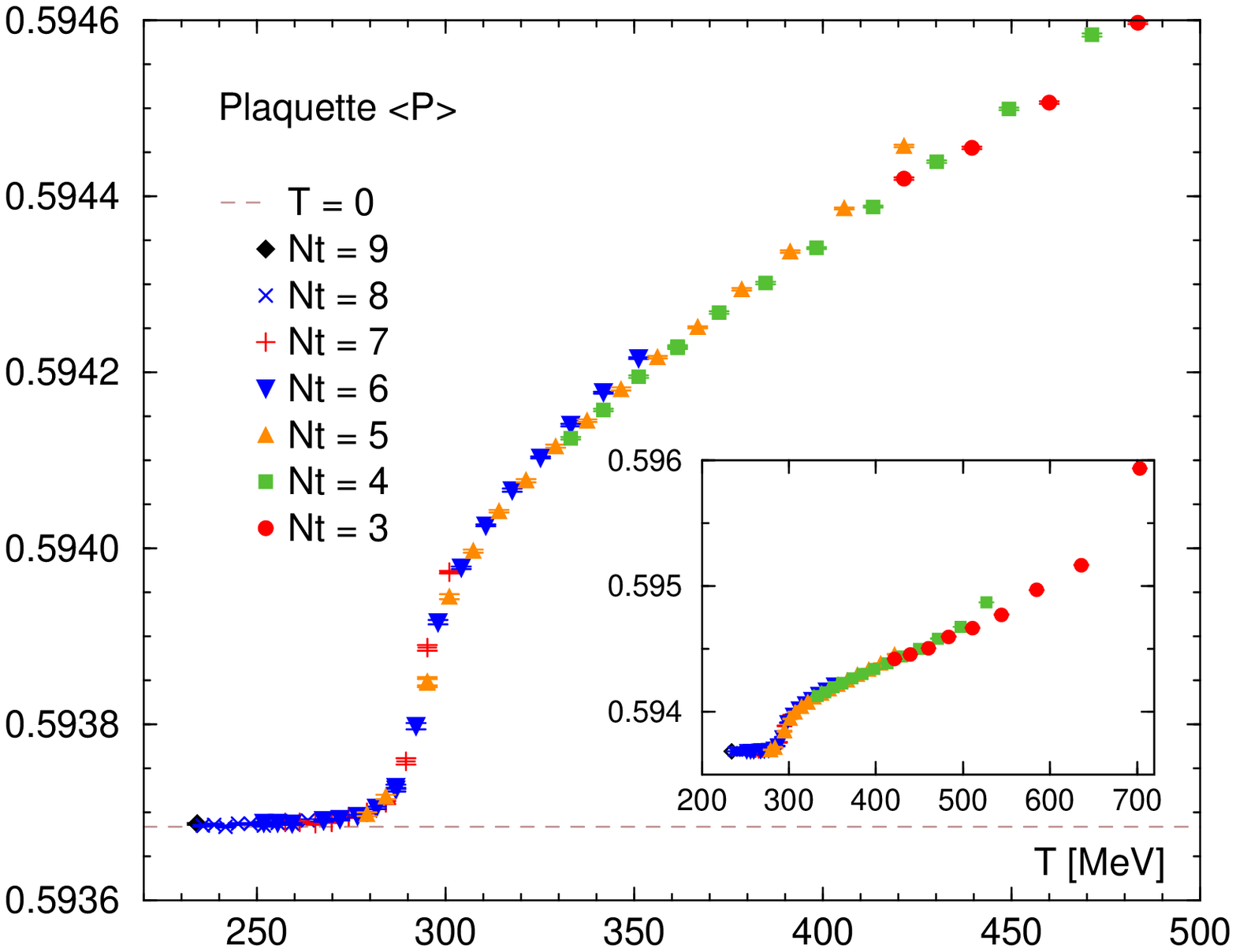}}
\caption{Plaquette expectation values plotted as a function of the physical temperature. $T=0$ value is measured on a symmetric $32^4$ lattice. The inset shows the same data over an expanded range.
}
\label{fig:plaq}
\end{figure}

In this section, present the lattice calculations of thermodynamics for the fixed-scale approach with the shifted boundary conditions. 
First, we plot the temperature dependence of the plaquette expectation values in Fig. \ref{fig:plaq}.
In contrast to the conventional fixed-scale approach, we can significantly increase the number of possible temperatures by using the shifted boundary conditions.
Although the plaquette values show smooth temperature dependence, small but significant deviations can be observed at higher temperatures.
We discuss these deviations in detail in a later section of the paper.

\subsection{Trace anomaly}
\label{sec:e3p}
In this subsection, we calculate the trace anomaly that is defined for the case with the shifted boundary conditions as the following.
\begin{eqnarray}
\frac{\epsilon-3p}{T^4} &=& \frac{1}{VT^3} a\frac{d\beta}{da} \left\langle 
\frac{dS_g}{d\beta} \right\rangle_{sub} \\
&=& 6 \left(N_t^2+\vec{s}^2\right)^2 a\frac{d\beta}{da} \langle P \rangle_{sub}.
\end{eqnarray}
Here, $a\frac{d\beta}{da}$ denotes the beta function , $V$ denotes the spatial volume and $\langle P \rangle_{sub}$ indicates the plaquette values, but with the zero-temperature value subtracted. 

By using the T-integral method \cite{Umeda:2008bd}, we can calculate the EOS, e.g., pressure, energy density, and entropy density, based on the trace anomaly. 
Therefore, the trace anomaly is the most basic quantity for the calculation of the EOS.
Here, we only discuss the trace anomaly, and other EOS quantities will be discussed in our forthcoming paper.

\begin{figure}[tb]
\resizebox{75mm}{58mm}{
\includegraphics{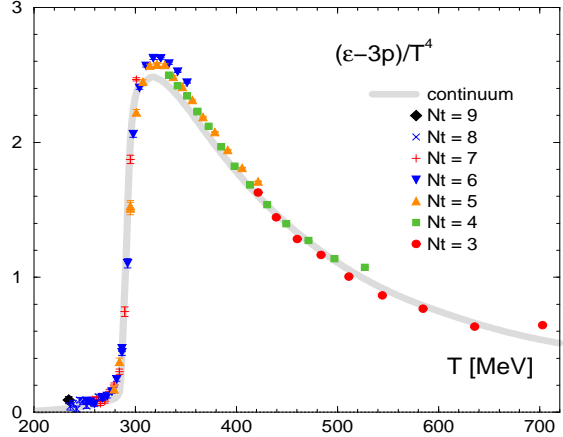}}
\caption{Trace anomaly as a function of temperature. The shifted boundary conditions aid in realizing the various temperature simulations at each $N_t$ at the fixed scale. The ``continuum'' line shows the data from Ref.~\cite{Borsanyi:2012ve}. 
}
\label{fig:e3p}
\end{figure}

\begin{figure}[tb]
\resizebox{75mm}{58mm}{
\includegraphics{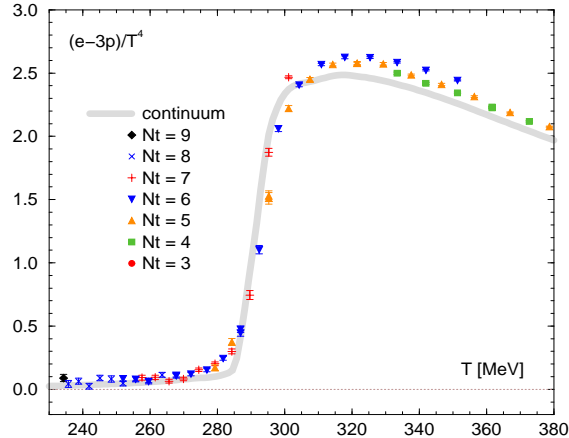}}
\caption{Trace anomaly as a function of temperature. The data are the same as those in Fig. \ref{fig:e3p}, but the temperature scale is magnified in the vicinity of the phase transition. 
}
\label{fig:e3p_low}
\end{figure}

Figure \ref{fig:e3p} shows the trace anomaly obtained at the single lattice scale of $\beta=6.0$ with the shifted boundary conditions.
Here, we adopted the beta function of $adg^{-2}/da=-0.098172$ \cite{Boyd:1996bx}.
In Figure \ref{fig:e3p} also includes the continuum limit curve of the trace anomaly as regards the SU(3) gauge theory \cite{Borsanyi:2012ve}, which is calculated in the conventional fixed $N_t$ approach.
Borsanyi et al. calculated the transition temperature based on the Polyakov loop susceptibility, and the temperatures are expressed as ratios with respect to the critical one. Here, we assume the critical temperature as $T_c=293$ MeV, which choice is discussed later, and show the continuum limit curve as a function of temperature in units of MeV. Since the continuum limit is not considered in our study, the continuum values are reference data.
Figure \ref{fig:e3p_low} also depicts the same data, but the temperature scale is magnified in the vicinity of the phase transition.

\begin{figure}[tb]
\resizebox{75mm}{58mm}{
\includegraphics{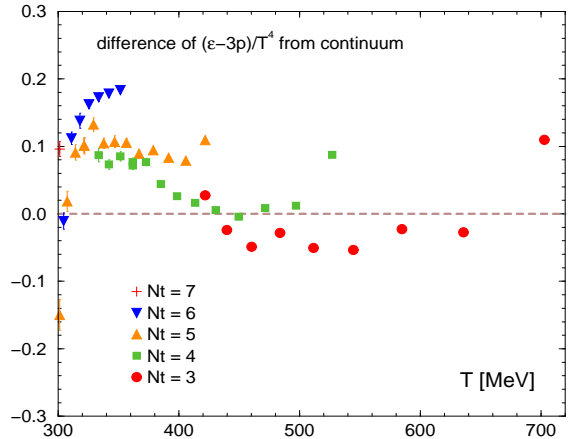}}
\caption{Difference between our trace anomaly results and the continuum values as a function of temperature. Statistical errors are nearly less than the size of the corresponding symbols.
}
\label{fig:e3pdiff}
\end{figure}

Our results are fairly consistent with the continuum values.
From Fig. \ref{fig:e3p_low}, we note that the simulations for a fixed value of $N_t$ with varying boundary shifts can correctly describe the trace anomaly both below and above $T_c$.
Upon examining the results in detail, we can observe visible differences between our results and the continuum values around the peak position.
In Ref.~\cite{Borsanyi:2012ve}, the authors have stated that cutoff effects are apparent around temperatures immediately above $T_c$, for example, the peak height at a fixed value of $N_t=6$ result is approximately 7\% greater than the continuum value. Therefore, the deviation appears to be reasonable upon considering the cutoff effects.
Furthermore, we also observe other deviations from the continuum values at the zero boundary shift for each $N_t$. 
In order to highlight the deviation, we replot our data as the difference from the continuum values in Fig. \ref{fig:e3pdiff}.
We can clearly observe that the zero-shift results deviate from the other values and continuum results.

We can understand the deviations by means of the non-interacting limit at finite lattice spacing, which was discussed with respect to Fig.~2 in Ref.~\cite{Giusti:2012yj}.
The figure shows pressure values divided by the Stefan--Boltzmann limit as a function of lattice cutoff in the case of the non-interacting limit.
Their result shows that shifted boundaries reduce the cutoff effects of EOS in the non-interacting limit when compared with that with zero boundary shift.
With this backdrop, we numerically confirmed that the shifted boundaries suppress cutoff effects even in the interacting case.

\subsection{Critical temperature}
\label{sec:tc}
In this subsection, we discuss the critical temperature. In the conventional fixed-scale approach, it is difficult to determine the critical temperature with good precision because of lower temperature resolution. 
As discussed thus far, the shifted boundary condition can be used to possibly overcome this issue.
In the SU(3) gauge theory, in general, the critical temperature is determined by the Polyakov loop expectation value and its susceptibility. 
However, when we adopt the shifted boundary conditions, the Polyakov loop defined along the temporal direction is no longer the correct order parameter of the transition because the compact direction is at an angle with respect to the temporal direction.
Although it is possible to define a gauge-invariant loop in the temporal direction with spatial hopping corresponding to the boundary shifts, this loop is merely a Polyakov loop in a moving frame.
If we use light quarks to construct the dressed Polyakov loop \cite{Bilgici:2008qy}, it may be possible to define the transition correctly. The analysis of this process will form our future work.

\begin{figure}[tb]
\resizebox{75mm}{58mm}{
\includegraphics{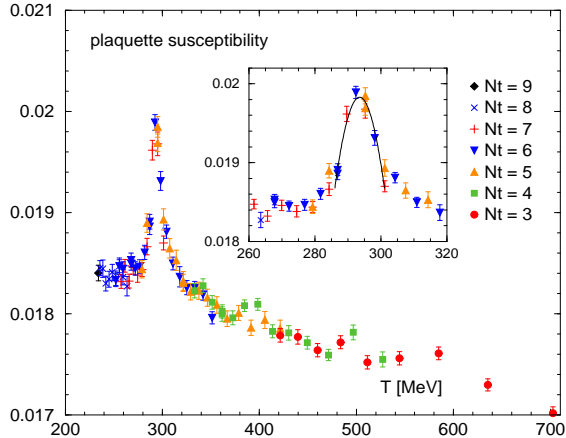}}
\caption{Plot of plaquette susceptibility as a function of temperature. A susceptibility peak is observed at a temperature of approximately 293 MeV, which value is obtained by a quadratic fit with the data of nine points closest to the peak. 
}
\label{fig:plaqsus}
\end{figure}

In order to determine the transition in this study, we utilize the plaquette susceptibility $\chi_P$ instead of the Polyakov loop,
\begin{eqnarray}
\chi_P = 6 N_s^3 N_t \left(  \langle P^2 \rangle - \langle P \rangle^2 \right).
\end{eqnarray}
Figure \ref{fig:plaqsus} shows the plaquette susceptibility as a function of temperature.
We note that the plaquette susceptibility exhibits a clear peak around the expected critical temperature, which is approximately 293 MeV as obtained by a quadratic fit using the data of nine points closest to the peak. 

\section{Beta function}
\label{sec:beta}
In the fixed-scale approach, the beta functions do not change with temperature. Although this is one of the advantages of the fixed-scale approach, the calculation of the beta function is still an obstacle to reduce zero-temperature computations.    
The beta functions are obtained from the coupling-parameter dependence of the lattice scale. To estimate the dependence by the method of finite differences with respect to the coupling parameters, additional simulations are necessary at zero temperature. 

Here, we comment on a method to calculate the beta functions without using additional zero-temperature simulations by using the shifted boundary conditions.
In this light, Giusti and Meyer have previously proposed the method to calculate thermodynamic quantities by using the shifted boundary conditions \cite{Giusti:2010bb}. In particular, with this method the entropy density can be derived without any zero-temperature simulation. 
On the other hand, in the fixed-scale approach, we can also calculate the entropy density as a function of temperature, but the beta function as an overall factor remains to be determined.
Therefore, the beta function can be determined by matching of the entropy densities at a temperature obtained by the shifted boundary and the T-integral method in the fixed-scale approach. 

The calculation of the entropy density by using the shifted boundary conditions can be time consuming. When the matching temperature is high, i.e., the temporal lattice extent is small, the numerical cost may be moderate. However, higher temperatures cause lattice artifacts in the EOS due to the UV temperature effect \cite{Umeda:2008bd}. Consequently, it is necessary to choose a moderate temperature for the matching.

Recently, another new method has been proposed to calculate the entropy density without zero-temperature simulation by using the gradient flow \cite{Suzuki:2013gza,Asakawa:2013laa}.  
Consequently, it is possible to choose the method for the matching of the beta function.

\section{Conclusion}
\label{sec:conclusion}
In this study, we attempt to investigate the EOS and the critical temperature specified by the SU(3) gauge theory by using the shifted boundary conditions in the fixed-scale approach. 
The shifted boundary condition can significantly increase the number of possible temperatures while retaining the advantages of the fixed-scale approach.
The trace anomaly obtained with the shifted boundary conditions is fairly accurate. When we adopt a sufficiently large spatial volume, calculations over a wide range of shift vectors yield fairly reasonable results, e.g., calculations with different $N_t$ at a fixed same temperature are consistent with each other.
Furthermore, we numerically confirmed that the shifted boundary conditions suppress lattice artifacts of the EOS. This suppression has previously been confirmed for the case of non-interacting limit \cite{Giusti:2012yj}.  

We also determined the critical temperature by using the plaquette susceptibility, which exhibits a clear peak at the expected temperature. This calculation is fairly complex when using the conventional fixed-scale approach. Further, problems arise regarding the definition of the Polyakov loop with the shifted boundary condition. The dressed Polyakov loop \cite{Bilgici:2008qy} can instead yield a valid order parameter even with the shifted boundary conditions.

Recently, certain new methods have been proposed to calculate thermodynamic quantities by means of the shifted boundary conditions \cite{Giusti:2010bb} and the gradient flow \cite{Suzuki:2013gza,Asakawa:2013laa}, which enable us to calculate entropy density using only finite-temperature configurations. 
These methods can also provide a new approach to calculate the beta function in combination with our T-integral method for calculating the entropy density. 
Further, the recently proposed methods enable us to calculate the pressure and the energy density, but the corresponding zero-temperature values are required even with those methods. In this light, our approach is worthy of further investigation in the study of QCD thermodynamics.   

\section*{Acknowledgments}
The author thanks A.~Nakamura, H.~Fukaya, H.~Matsufuru, K.~Kanaya, and S.~Ejiri for helpful discussions and comments. This work was supported by JSPS KAKENHI Grant Numbers 22740168 and 26400251. The numerical calculations were carried out on the SX-8R system at RCNP in Osaka University and the SR16000 system at the High Energy Accelerator Research Organization (KEK). 
This work is also partly supported by the Large Scale Simulation Program of KEK No. 13/14-21.

\end{document}